\documentclass[a4paper,epsfig,12pt]{article}
\usepackage[pctex32]{graphics}

\newcommand{\sect}[1]{\setcounter{equation}{0}\section{#1}}

\begin{document}
\topmargin 0pt \oddsidemargin 0mm

\renewcommand{\thefootnote}{\fnsymbol{footnote}}
\begin{titlepage}
\begin{flushright}
INJE-TP-03-01\\
hep-th/0301073
\end{flushright}

\vspace{5mm}
\begin{center}
{\Large \bf Holographic entropy bounds in the inflationary universe} \vspace{12mm}

{\large  Yun Soo Myung\footnote{e-mail
 address: ysmyung@physics.inje.ac.kr}}
 \\
\vspace{10mm} {\em  Relativity Research Center and School of Computer Aided
Science, Inje University Gimhae 621-749, Korea}
\end{center}

\vspace{5mm} \centerline{{\bf{Abstract}}}
 \vspace{5mm}
We introduce the relation between the holographic entropy bounds and the inflationary universe.
First  the holographic entropy bounds for radiation-dominated universe,
radiation-dominated universe with a positive cosmological constant are introduced. For  an
exact de Sitter phase, we  use the maximal entropy bound.
We classify  the inflation based on the quasi-de Sitter spacetime into
three steps: slow-roll period of inflation,   epoch of
reheating, and  radiation-dominated era. Then  we study how to apply
three entropy bounds to the three steps of the inflation.
 Finally we discuss our results.

\end{titlepage}

\newpage
\renewcommand{\thefootnote}{\arabic{footnote}}
\setcounter{footnote}{0} \setcounter{page}{2}

\sect{Introduction}

The inflation turned out to be a successful tool to
resolve the problems of the  hot big bang model~\cite{Infl}. Thanks to the
recent observations of the cosmic microwave background
anisotropies and large scale structure galaxy surveys, it has
become widely accepted by the cosmology community~\cite{JGB}.
The idea of inflation is based on the very early universe
dominance of vacuum energy density of a hypothetical scalar field, the
inflaton. This produces the quasi-de Sitter spacetime~\cite{Hogan} and during
the slow-roll period, the equation of state can be approximated by
the vacuum state as $p\approx -\rho$~\cite{FKo}. After that there must exist
a strong non-adiabatic and out-of-equilibrium phase called
reheating to produce a large increase of the entropy. But we don't
know exactly how inflation started.

To solve this problem we have
to build cosmology from the quantum gravity, but now  we are far from
it. Although we are lacking for a complete understanding of the
quantum gravity, there exists the holographic principle.
 This principle  is mainly based on  the idea that for a given volume $V$, the state
 of maximal entropy is given by the largest black hole that fits inside $V$.
 't Hooft and Susskind~\cite{Hooft} argued that the microscopic entropy $S$
 associated with the volume $V$ should be less than the  Bekenstein-Hawking
entropy:  $S \le A/4G$ in the units of $c=\hbar=1$~\cite{Beke}. Here the horizon area $A$ of a
black hole equals the surface area of the boundary of $V$.
That is,  if one reconciles quantum mechanics and
gravity, the observable degrees of freedom of the three-dimensional universe
comes from a two-dimensional surface. Actually holographic area
bounds limit  the number of physical degrees of freedom in the
bulk spacetime.

The implications of the holographic principle  for cosmology have
been investigated in the  literature.
Following an earlier work by Fischler and Susskind~\cite{FS} and works
in~\cite{Hubb,Bous}, it was argued  that the maximal entropy inside the
universe is produced by the Hubble horizon-size black hole.
Roughly  the total entropy should be
less or equal than the Bekenstein-Hawking entropy of the
Hubble-size black hole ($\approx H V_{\rm H}/4G_{n+1}$) times the number
($N_{\rm H}\approx V/V_{\rm H}$) of Hubble regions in the
universe. Hence one obtains an upper bound on the total entropy
which is proportional to  $HV/4G_{n+1}$.

Furthermore, Verlinde obtained the pre-factor as $(n-1)$ and
 proposed the new holographic bound like Eq.(\ref{2eq6}) in
a radiation-dominated phase by introducing three entropies~\cite{Verl} : Bekenstein-Verlinde
entropy ($S_{\rm BV}$), Bekenstein-Hawking
entropy ($S_{\rm BH}$), and Hubble
entropy ($S_{\rm H}$).
 For example, such a radiation is
given by a conformal field theory (CFT) with a large central charge dual
to the AdS-black hole~\cite{SV}. It indicated an interesting relationship
between the Friedmann equation governing the cosmological evolution and
the square-root form of entropy-energy relation, called
Cardy-Verlinde formula~\cite{Cardy}. Although the Friedmann equation has a geometric
origin and  the Cardy-Verlinde formula is designed only  for the matter
content, this strongly suggested  that both sets  may arise
from a single underlying fundamental theory. In addition, this
approach  showed new features about the mathematical structure of
the Friedmann equation~\cite{AAC}.

In this work we will explore the implications of the holographic principle for  the
inflationary universe. We classify  the inflation based on the quasi-de Sitter spacetime into
three steps : (i) slow-roll period of inflation (SR), (ii)  epoch of
reheating (RH), and (iii) radiation-dominated era (RD).
 In order to obtain  the holographic entropy bounds for these steps,
 we introduce three kinds of the universe with the  equation of state:
 an  exact de Sitter phase with a positive cosmological constant,
 a radiation-dominated universe   with a positive cosmological
 constant,
  and a radiation-dominated universe.
 In order to get  the entropy bounds for a radiation-dominated universe
  with a positive cosmological constant, we need to define the
  entropy for the cosmological constant ($S_{\rm \Lambda})$ and
  the cosmological D-entropy ($S_{\rm D}$)~\cite{CM1}. Furthermore, we
  use the entropy bounds for  de Sitter space.

The organization of this paper is as follows. In section 2, we
discuss the holographic entropy bounds for radiation-dominated,
radiation-dominated with a positive cosmological constant.
 Section 3 is devoted to the description of  the inflationary
universe based on the quasi-de Sitter spacetime. And then we study how to apply
the entropy bounds to the three steps of the inflation: SR, RH,
and RD. Finally we discuss our results in section 4.

\section{Cosmological entropy bounds}
\subsection{Three entropies}
Let us start a $(n+1)$-dimensional Friedmann-Robertson-Walker (FRW)
mertic
\begin{equation}
\label{2eq1} ds^2 =-d\tau^2 +R(\tau)^2 \Big[ \frac{dr^2}{1-kr^2} +r^2 d\Omega^2_{n-1} \Big],
\end{equation}
where $R$ is the  scale factor of the universe and $d\Omega^2_{n-1}$
denotes the line element of a $(n-1)$-dimensional unit sphere.
Here $k=-1,~0,~1$ represent that  the universe  is
open, flat, closed, respectively.
The evolution is determined by the two FRW equations
\begin{eqnarray}
\label{2eq2}
 && H^2 =\frac{16\pi G_{n+1}}{n(n-1)}\frac{E}{V}
-\frac{k}{R^2}
     +\frac{1}{l^2_{n+1}}, \nonumber \\
&& \dot H =-\frac{8\pi G_{n+1}}{n-1}\left (\frac{E}{V} +p\right)
    +\frac{k}{R^2},
\end{eqnarray}
where $H$ represents the Hubble parameter with the definition
$H=\dot R/R$ and the overdot stands for  derivative with
respect to the cosmic time $\tau$,  $E$ is the energy of matter
filling the universe, and $p$ is its pressure. $V$ is the
volume of the universe, $V=R^n \Sigma^n_k$ with $\Sigma^n_k$ being the
volume of a $n$-dimensional space with $k$, and $G_{n+1}$ is the
Newton constant in ($n+1$) dimensions. Here we assume the equation of state for matter:
 $p=\omega \rho,~ \rho=E/V$. For our purpose, we include
 the curvature scale of de Sitter space $l_{n+1}$  which relates to the cosmological constant
 via $1/l^2_{n+1}=2\Lambda_{n+1}/n(n-1)$.
 Youm~\cite{Youm1} considered the cosmological constant as the energy density
 $p_{\rm \Lambda}=-\rho_{\rm \Lambda}=-\Lambda_{n+1}$ to obtain
 the entropy bounds for  a vacuum-like matter with $\omega=-1$. Hereafter we do
 not follow this direction to obtain the entropy bound for a positive cosmological constant.

 Verlinde has introduced three entropies
 for a closed radiation-dominated universe~\cite{Verl}\footnote{In \cite{Verl} the first one is called the
Bekenstein entropy. In fact this bound is slightly different from
the original Bekenstein entropy~\cite{Beke} by a numerical factor
$1/n$. So we call this the Bekenstein-Verlinde entropy. This
could be viewed as the counterpart of the Bekenstein entropy in the
cosmological setting~\cite{CMO}.}:
\begin{eqnarray}
\label{2eq3}
 {\rm Bekenstein-Verlinde\ entropy}:&& S_{\rm BV}=\frac{2\pi}{n}ER
   \nonumber \\
 {\rm Bekenstein-Hawking\ entropy}:&& S_{\rm BH}=(n-1)\frac{V}{4G_{n+1}R}
    \nonumber \\
  {\rm Hubble\ entropy}:&& S_{\rm H}=(n-1)\frac{HV}{4G_{n+1}}.
\end{eqnarray}
 $S_{\rm BV} \le S_{\rm BH}$ is supposed to hold for a weakly
self-gravitating universe ($HR \le 1$), while
 $S_{\rm BV} \ge S_{\rm BH}$ works when the universe is in the strongly self-gravitating
phase ($HR \ge 1$).
 It is interesting to
note that for  $k=1,HR=1$, one finds that three entropies are identical:
 $S_{\rm BV}= S_{\rm BH}=S_{\rm H}$.


\subsection{Radiation-dominated universe}

First we start with $k=1,\Lambda_{n+1}=0$ case because this case gives us a
concrete and clear relation.
We  define a quantity $E_{\rm BH}$ which corresponds to energy
needed to form a universe-size black hole :
$ S_{\rm BH}=(n-1)V/4G_{n+1}R \equiv 2\pi E_{\rm BH}
R/n $. With this quantity, the Friedmann equations (\ref{2eq2}) can
be further cast to
\begin{eqnarray}
\label{2eq4}
 && S_{\rm H}=\frac{2\pi R}{n}\sqrt{E_{\rm
BH}(2E-E_{\rm
BH})}, \nonumber \\
&& E_{\rm BH}=n(E+pV -T_{\rm H} S_{\rm H}),
\end{eqnarray}
where the Hubble temperature ($T_{\rm H}$) is given by  $T_{\rm H}=-\frac{\dot H}{ 2\pi H}$.
On the other hand,  the  entropy  of
radiation  and its Casimir energy   can be described  by the
Cardy-Verlinde  and Smarr formulae
\begin{eqnarray}
\label{2eq5}
&& S =\frac{2\pi R}{n}\sqrt{E_c(2E-E_c)},
  \nonumber \\
&& E_c=n(E+pV -T S).
\end{eqnarray}
Here $T$ stands for the temperature of  radiation with $\omega=1/3$.
These  can describe the entropy $S$ of a CFT-radiation
living on a $n$-dimensional sphere with radius $R$. Here $E$ is
the total energy of the CFT and $E_c$ stands for the Casimir
energy of the system,  non-extensive part of the total energy.
Suppose the entropy of  radiation  in the FRW
universe can be described by the Cardy-Verlinde formula. Comparing
(\ref{2eq4}) with (\ref{2eq5}), one  finds that if
$E_{\rm BH}=E_c$, then $S_{\rm H}=S$ and $T_{\rm H}=T$.
At this stage we  introduce the Hubble  bounds for entropy,
temperature and Casimir energy~\cite{Verl}
 \begin{equation}
 \label{2eq6}
 S \le S_{\rm H},~~ T \ge T_{\rm H},~~~E_c \le E_{\rm BH}, ~~{\rm for}~ HR \ge 1
 \end{equation}
which are relation between
geometric and matter quantities.
The Hubble entropy bound is saturated by the entropy
of radiation filling  the universe if the Casimir energy $E_c$ is
 enough to form a universe-size black hole.
At this moment, equations (\ref{2eq4}) and (\ref{2eq5}) coincide exactly.
  This implies that the first Friedmann equation
somehow knows the entropy formula of a square-root form for radiation-matter filling the
universe. For example, let us  consider a moving brane universe in the
background of the 5D Schwarzschild-AdS black hole. Savonije and
Verlinde~\cite{SV} found that when the brane crosses the black
hole horizon, the Hubble entropy bound  is saturated by the entropy
of black hole(=the entropy of the  CFT-radiation). Also the Hubble temperature and energy ($T_{\rm H},E_{\rm BH}$)
equals to the temperature and Casimir energy ($T,E_c$)
of the CFT-radiation dual to the AdS black
hole at this moment.

Up to now we discuss $k=1$ case only.
For later purpose, we list  the Hubble entropy bound for arbitrary $k$~\cite{Youm2}
 \begin{equation}
 \label{2eq7}
 S \le S_{\rm H}, ~~{\rm for}~ HR \ge  \sqrt{2-k}.
 \end{equation}

On the other hand,  the Bekebstein-Verlinde entropy bound for  arbitrary $k$ is given by
 \begin{equation}
 \label{2eq8}
 S \le S_{\rm BV}, ~~{\rm for}~ HR \le  \sqrt{2-k}.
 \end{equation}

\subsection{Radiation-dominated universe  with a
positive cosmological constant }

For a radiation-dominated  FRW universe with $k=1,\Lambda_{n+1} \not=0$,
we have to introduce the cosmological entropy
$S_{\rm \Lambda}$,
 cosmological D-entropy $S_{\rm D}$  and D-temperature $T_{\rm D}$
 as~\cite{CM1}
 \footnote{Although Bousso argued that
a cosmological constant did not carry entropy~\cite{Bous2},
 there is no contradiction to introducing
the corresponding entropy. Actually $S_{\rm \Lambda}$ is closely related to the de Sitter
entropy of $S_{\rm dS}$. This is given by the Bekenstein-Hawking entropy of the
de Sitter horizon ($(n-1)  V_{\rm dS}/4G_{n+1} l_{n+1} \approx S_{\rm dS}$) times the number
($N_{\rm dS}= V/V_{\rm dS}$) of de Sitter regions in the
universe. }
\begin{equation}
\label{2eq9}
S_{\rm \Lambda}=(n-1) \frac{V}{4 G_{n+1}l_{n+1}}, ~~
S_{\rm D} =\sqrt{|S^2_{\rm H}-S^2_{\rm \Lambda}|}
,~~T_{\rm D}=- \frac{ \dot H} { 2 \pi \sqrt{|1/l^2-H^2|}}.
\end{equation}

We note that the cosmological D-entropy $S_{\rm D}$ is constructed
by analogy with the difference (D) between the entropy of exact de
Sitter space and that of asymptotically de Sitter space. Further
we assume that three entropies in Eq.(\ref{2eq3}) are still useful
for describing the radiation-dominated universe  with $\Lambda_{n+1}
\not=0$. In the case of $\Lambda_{n+1} \to 0$, one recovers
the radiation-dominated universe without a cosmological constant.
\begin{equation}
\label{2eq10}
S_{\rm \Lambda} \to 0, ~~
S_{\rm D} \to S_{\rm H}
,~~T_{\rm D} \to T_{\rm H}.
\end{equation}
For $S_{\rm H} \ge S_{\Lambda}$, the Friedmann equations in Eq.(\ref{2eq1}) can be
rewritten as
\begin{eqnarray}
\label{2eq11}
 && S_{\rm D}=\frac{2\pi R}{n}\sqrt{E_{\rm
BH}(2E-E_{\rm
BH})}, \nonumber \\
&& E_{\rm BH}=n(E+pV -T_{\rm D} S_{\rm D}),
\end{eqnarray}
 while the entropy and  Casimir energy  of
radiation can be expressed as
\begin{eqnarray}
\label{2eq12}
&& S =\frac{2\pi R}{n}\sqrt{E_c(2E-E_c)},
  \nonumber \\
&& E_c=n(E+pV -T S).
\end{eqnarray}
On the other hand, for $S_{\rm H} \le S_{\rm \Lambda}$, the
 equations can be rewritten as
\begin{eqnarray}
\label{2eq13}
 && S_{\rm D}=\frac{2\pi R}{n}\sqrt{E_{\rm
BH}(E_{\rm
BH}-2E )}, \nonumber \\
&& E_{\rm BH}=n(E+pV -T_{\rm D} S_{\rm D})
\end{eqnarray}
and the entropy and Casimir energy  are
\begin{eqnarray}
\label{2eq14}
 && S =\frac{2\pi R}{n}\sqrt{E_c(E_c-2E)},
  \nonumber \\
&& E_c=n(E+pV -T S).
\end{eqnarray}
As is shown in Eq.(\ref{2eq10}),
the cosmological
D-entropy
plays the same role as the Hubble entropy does in the case without cosmological
constant.

Now we are in a position to discuss how the entropy bounds are changed.
The first Friedmann equation can be rewritten as
\begin{equation}
\label{2eq15}
(HR)^2-\frac{R^2}{l_{n+1}^2}= 2 \frac{S_{\rm BV}}{S_{\rm BH}} -k.
\end{equation}
Using this relation, in  case of $k=1, \Lambda_{n+1}=0$,
one finds $HR \ge 1 \to S_{\rm BV} \ge
S_{\rm BH}$, while $HR \le 1 \to S_{\rm BV} \le
S_{\rm BH}$. Hence this leads to the Hubble entropy bound of $S \le S_{\rm H}$ for $HR \ge 1$,
whereas the Bekenstein-Verlinde entropy bound of  $S \le S_{\rm BV}$ for $HR \le 1$.
Other cases of $k=-1,0$ are shown explicitly in Eqs.(\ref{2eq7}) and
(\ref{2eq8}).

For  $k=-1,0,1$ and $ \Lambda_{n+1} \not=0$, $(HR)^2 -\frac{R^2}{l_{n+1}^2} \ge 2-k
  \to S_{\rm BV} \ge
S_{\rm BH}$, while $ (HR)^2 - \frac{R^2}{l_{n+1}^2}\le 2-k  \to S_{\rm BV} \le
S_{\rm BH}$.
Thus  this leads to the D-entropy bound for the strongly self-gravitating universe:
\begin{equation}
\label{2eq16}
S \le S_{\rm D},~~ {\rm for}~~
HR \ge \sqrt {2-k+ \frac{R^2}{l_{n+1}^2}},
\end{equation}
whereas the Bekenstein-Verlinde entropy bound is found for the weakly self-gravitating
system:
\begin{equation}
\label{2eq17}
S \le S_{\rm BV},~~ {\rm for}~~
HR \le \sqrt {2-k+ \frac{R^2}{l_{n+1}^2}}.
\end{equation}
When the  D-entropy bound is saturated by the
entropy $S$ of radiation, both sets of equations (\ref{2eq11})
[or (\ref{2eq13})] and (\ref{2eq12}) [or (\ref{2eq14})] coincide with
each other, just like the case without the cosmological constant.
We note from Eq.(\ref{2eq15}) that one cannot find the relation of $S_{\rm D}=
S_{\rm BV}=S_{\rm BH}$ for $HR=1$, unless $\Lambda_{n+1}=0,k=1$.
Here we obtain an important relation for $k=0$ case from Eq.(\ref{2eq15})
as
\begin{equation}
\label{2eq18}
Hl_{n+1} \ge 1.
\end{equation}
This relation comes out because three entropies all in  Eq.(\ref{2eq3})
should be positive definitely. Applying this to the entropy,
we have
\begin{equation}
\label{2eq19}
S_{\rm H} \ge S_{\rm \Lambda}.
\end{equation}
 Hence the other case of
$S_{\rm H} \le S_{\rm \Lambda}$ in Eqs.(\ref{2eq13}) together with (\ref{2eq14})
is not allowed for flat ($k=0$) cosmological evolution.
 In other words, this case is forbidden due to the evolution
 equation.

\subsection{ de Sitter Universe  with a positive cosmological constant}

Now we wish to discuss the entropy bounds for the exact de Sitter
evolution with $\Lambda_{n+1}$. This corresponds to the perfect
fluid $p=- \rho$ with $ \rho=\Lambda_{n+1}$.  We need this case because
the equation of state for  slow-roll
period of inflation is not defined clearly but it is approximately given by $p\approx-\rho$.
 In this case it is easily conjectured that we cannot
obtain the desired entropy bounds
\footnote{However, assuming the adiabatic expansion of the FRW universe, one has the
lower entropy bound for the de Sitter expansion for $k=1$~\cite{Youm1}
$S \ge S_0 \Big( \frac{R}{H} \Big)^n$
for $HR \ge1$, while one finds the constant lower bound
$S \ge \tilde S_0 \Big( \frac{1}{\Lambda_{n+1}} \Big)^n$
for $HR \le 1$. Here  we do not follow this approach.} because the first term of the
right hand side of Eq.(\ref{2eq15}) is absent. In other words, the
first Friedmann equation does not play an important  role in determining the
corresponding entropy bounds. What we can obtain at most here is that the cosmological entropy
 $S_{\rm \Lambda}$ is equal to the Hubble entropy $S_{\rm H}$.
In other words, Eq.(\ref{2eq15}) with $k=0$ shows that in a universe dominated by the
cosmological constant, the solution is an exponential expansion of
rate of $R(\tau)\propto e^{H \tau}=e^{\tau/l_{n+1}}$.

\sect{Inflationary Universe and Entropy Bounds}

\subsection{Inflationary Universe on the quasi-de Sitter space}

We adopt  an idealized model of the inflationary model based on
the quasi-de Sitter spacetime~\cite{Hogan}. In what follows we work with the
(3+1)-dimensional flat FRW slicing of de Sitter spacetime, because
 this maps directly onto the FRW spacetime of  the
 post inflationary universe. The line element which covers half
 of the full de Sitter solution is given by
\begin{equation}
\label{3eq1}
ds^2_{FRW-dS}=-d\tau^2 + \exp[2H\tau]\Big(dr^2 +r^2 d\Omega_2^2 \Big).
\end{equation}
Another slicing of an exact de Sitter spacetime is given by the static
coordinates\footnote{ The coordinates of $\tau,r$ and $t,\tilde r$
are related by the transformations $r=e^{H t} \frac{\tilde r}{\sqrt{ 1-H^2\tilde r^2}},
\tau= t +\frac{1}{2H} \ln[1-H^2 \tilde r^2]$~\cite{FKo}.}
\begin{equation}
\label{3eq2}
ds^2_{S-dS}=-\Big(1-H^2 \tilde r^2\Big)dt^2  +
\Big(1-H^2 \tilde r^2 \Big)^{-1}d \tilde r^2 + \tilde r^2d\Omega^2_2,
\end{equation}
where $H^{-1}=l_4$ is the size of the event (cosmological) horizon.
The Gibbons-Hawking temperature is  $T_{\rm GH}=H/2 \pi$
~\cite{GH} and the area of the event horizon is $A=
4\pi/H^2$. These coordinates cover the entire causal diamond
accessible to any actual observation by an observer at the origin
of $r=0$. Hence this finite spatial region is  subjected to
the holographic entropy bound of $S \le S_{\rm dS}=A/4G=\pi/GH^2=\pi l^2_4/G$.
 We call this a ``hot tin can" in the
sense that an observer in the causal patch (the interior of the
can) is surrounded by a hot horizon (the walls)~\cite{Hogan}. The physical
degrees of freedom, accessible to the observer, are in thermal
equilibrium with the Gibbons-Hawking temperature. Apparently an
inflating region resembles this exact de Sitter space, where the
apparent horizon and the event horizon coincide and thus there is
no exterior of the can.
However an actual situation is slightly different. Unlike the exact de Sitter space, during
inflation many modes are expelled from the apparent horizon. In
this case, the  space is not an exact de Sitter one (hat tin can) but a
quasi-de Sitter space (hot porous tin can). Authors in ref.~\cite{AKS} accounted this leaking
entropy $S_{\rm L}$ from the apparent horizon
to obtain a holographic limitation of the effective field
theory for inflation.

In order to show the route of information flow from inflation to
observable anisotropy, let us see the Penrose diagram in Fig.1.

\begin{figure}[here]
\hfil\scalebox{0.95}{\includegraphics{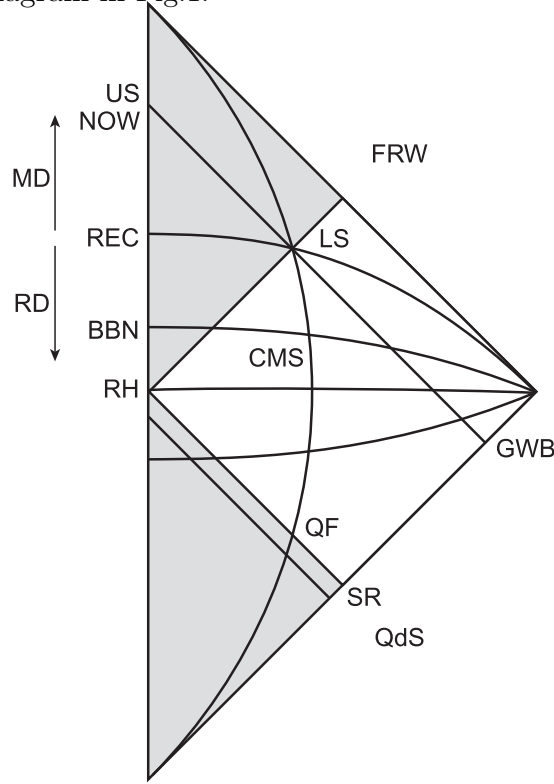}}\hfil
\caption{Penrose diagram of an inflationary cosmology based on  quasi-de Sitter space
and Friedmann-Robertson-Walker space}
\end{figure}

As usual, points denote two-spheres($S^2$), the left-hand edge
represents the world line of an observer at the origin. Others
are boundaries at infinity. The lower half stands for a quasi-de
Sitter space (QdS) and the upper half for  a FRW spacetime. The join
between them is the epoch of reheating (RH) and shaded regions of each show
the regions within the apparent horizon of an observer at the origin : one is
an apparent event horizon for QdS and
 the other is an apparent particle horizon for FRW.
 REC (BBN) represent spacelike hypersurfaces for the
recombination epoch and big bang nucleosynthesis epoch. Two regions in
QdS are necessary, one appropriate for matching onto FRW and the
other for holographic analysis. Actually perturbations are
imprinted by fluctuating quantum field (QF) on the scale of the
apparent event horizon  during the slow-roll period of
inflation (SR). The apparent horizon grows slightly during SR, as
is shown by two closely parallel null lines. This happens  because the spacetime becomes
asymptotically de Sitter space due to the increase  of entropy during
SR. The intersection of our past light cone (null line) with REC
is the two-sphere of the last scattering surface (LS) for the
cosmic background radiation. A particular timelike trajectory of a comoving
 sphere (CMS) is shown. The radiation-dominated era (RD) is from
 the end of RH to the time of REC and the matter-dominated era (MD) is
 extended from REC to the present: US, NOW. Finally a high frequency
 gravitational wave background (GWR) can reach US via direct null
 trajectories.

\subsection{Entropy bounds in the inflationary universe}

We begin with a scalar field ($\phi \equiv \phi(\tau)$ :
inflaton).
This gives us the energy density and pressure
\begin{equation}
\label{3eq3}
\rho_{\phi}=\frac{\dot \phi^2}{2} +V(\phi),~~
p_{\phi}=\frac{\dot \phi^2}{2} -V(\phi).
\end{equation}
Note that although the scalar field acts as a perfect fluid, it
does not possess any equation of state like $p_{\phi}=\omega_{\phi}
\rho_{\phi}$ exactly.
To get a good approximation during inflation, we consider only an
inflaton coupled to gravity minimally as
\begin{equation}
\label{3eq4}
3 H^2= \frac{\dot \phi^2}{2 M_p^2} +\frac{V(\phi)}{M_p^2},~~
\ddot \phi+ 3 H \dot \phi + \frac{\partial V(\phi)}{\partial \phi}=0,
\end{equation}
where the overdot is the time derivative and the Planck mass is given by
$M_p=1/\sqrt{8 \pi G_4}$
in the units of $c=\hbar=1$. The first equation is obtained from
Eqs.(\ref{2eq2}) and (\ref{3eq3}),
whereas the second from the conservation law of $\dot \rho
+3H(\rho+p)=0$.
Inflation occurs when the potential
energy of the scalar is dominant in Eq.(\ref{3eq3}). Then this situation is
approximated by the slow-roll period of inflation (SR). This is formally defined
by $|\epsilon| $ and $ |\delta|\ll 1$, where $\epsilon=\frac{3 \dot \phi^2}{2V},
 \delta=\frac{ \ddot \phi}{H\dot \phi}$. Actually the slow-roll
 approximation corresponds to dropping the terms of order ${\cal
 O}(\epsilon,\delta)$.
 Then the equation (\ref{3eq3}) leads to
\begin{equation}
\label{3eq5}
3 H^2 \approx \frac{V(\phi)}{M_p^2},~~
 3 H \dot \phi + \frac{\partial V(\phi)}{\partial \phi} \approx 0,
\end{equation}
Now let us discuss the evolution of physical scales as the universe
evolves.
The relevant picture is introduced in Fig.2 which may be acceptable in
the inflationary cosmology~\cite{LL}. 

\begin{figure}[here]
\hfil\scalebox{0.95}{\includegraphics{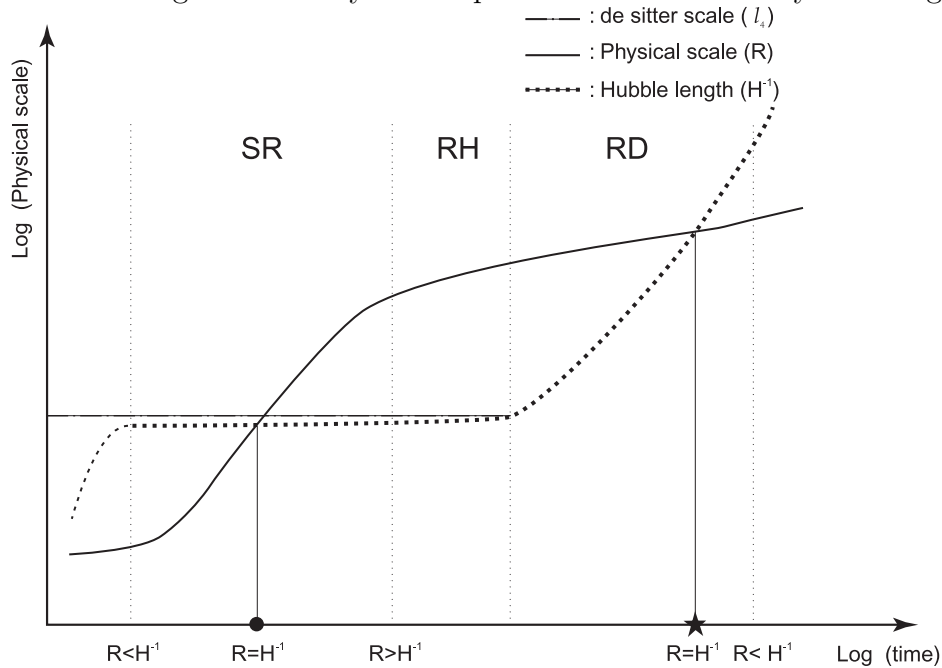}}\hfil
\caption{The behaviours of three scales. A physical scale ($R$) starts well inside the Hubble length
($H^{-1}$), then crosses outside sometime before the end of inflation, reentering long after inflation is over.
Also the de Sitter scale ($l_4$) is introduced. The vertical axis covers many powers of 10 in scale.}
\end{figure}
An important question concerning a given scale
is whether it is larger or smaller than the  Hubble horizon
 \footnote{Strictly speaking, we should refer to ``Hubble
distance" or ``Hubble length" for $H^{-1}$.}.  A physical scale $R$ starts
well inside the horizon and then crosses the
horizon  $R=H^{-1}$ at time ($\bullet$) during SR. It stays outside the horizon
during time : after  the inflation (SR), RH, and RD. It crosses the horizon $R=H^{-1}$ again
at  later time ($\star$) and reenters the horizon in the period of RD.
We wish to describe  the inflationary universe mainly in terms of
three periods: SR, RH, and RD.

(i) SR:
In the slow-roll approximation the potential and thus the  Hubble radius
of  $H^{-1} (\approx M_p/\sqrt{3V}$)
can be taken to be constant over each Hubble time during
inflation. Hence this can be approximated  by  de Sitter  universe with a
positive cosmological constant. Hence
the holographic entropy bound is that the observable entropy of
the universe cannot exceed the entropy of  exact de Sitter space\cite{Bous2},
 $S \le S_{\rm dS}=A/4G=\pi l^2_4/G \approx \pi M^2_p/3GV$.
This  is an absolute maximal entropy because of the nature of de
Sitter space. That is, the bound includes all degrees of freedom of
matter fields as well as all quantum degrees of freedom of the
spacetime itself. This is   the hat tin can-picture of de
Sitter space. There exist another picture of the hot porous tin
can where the total entropy in SR is divided into the dominant
equilibrium contribution approximately by the $S_{\rm dS}$
and a small non-equilibrium contribution by the leaking entropy
from the apparent horizon $S_{\rm L}$.

(ii) RH :
The slow-roll period  will cease as the
inflaton  moves to the minimum of the potential, where it
oscillates and decays into radiation, reheating the
inflated universe  to give  a large increase of the  entropy.
Although this is a non-adiabatic and out-of-equilibrium phase,
we may  approximate this phase by the radiation-dominated universe with a
positive cosmological constant. Here we still need a positive cosmological term
because during RH the universe  exhausts the vacuum energy. The promising entropy bound is
$S \le S_{\rm D}$ in Eq.(\ref{2eq16}) together with $S_{\rm H} \ge S_{\rm \Lambda}$
in Eq.(\ref{2eq19}).

(iii) RD :
 After  RH, the evolution of the universe is described by the
$k=0$ radiation-dominated one. Then the desired entropy bound is given
by $S \le S_{\rm H}$  in  Eq.(\ref{2eq7}) for the left hand side
of the reference point ($\star$)
and $S \le S_{\rm BV}$  in  Eq.(\ref{2eq8}) for the right hand side.

\section{Discussion}
We  discuss the relationship between three periods of the quasi-de Sitter spacetime
and three kinds of holographic entropy bounds.
Although this relation is not confirmed completely,
our approach  may be useful for   estimating the physical degrees of
freedom by the holographic principle.
Actually  the  $k=0$ case is more suitable for describing the
inflationary universe than $k=1$ because the $k$-curvature term  becomes less important
compared with the vacuum energy density term. However,  applying for   the holographic
entropy bounds to the inflation, $k=1$ case   is   more  intuitive than $k=0$.

 This is so
because $k=1$ case gives us the direct
 relations for the radiation-dominated universe: the Hubble bound of $S \le S_{\rm H}$
 is  valid for $R \ge H^{-1}$, while the Bekenstein-Verlinde  bound of $S \le S_{\rm BV}$
 is valid for  $R \le H^{-1}$. We note that the Hubble
 bound is based on how many  Hubble-size  black holes exist
 within  the radius $R$ of the universe.
 When the radius $R$ of the universe is smaller than the Hubble
 radius, one should reconsider the validity of the Hubble bound.
 In this case, the appropriate entropy bound is the
 Bekenstein-Verlinde bound. At the reference point ($\star$) of $R=H^{-1}$,
 one finds that $S_{\rm H}=S_{\rm BH}$, where $S_{\rm BH}$ is the Bekenstein-Hawking
 entropy of the universe-size black hole. Actually this is not to serve as a bound
 on the total entropy but rather on the sub-extensive component of the entropy for a finite
 system.  For $k=0$ case, the Hubble bound of $S \le S_{\rm H}$
 is  valid for $R \ge \sqrt{2} H^{-1}$,
 while the Bekenstein-Verlinde bound of $S \le S_{\rm BV}$
 is valid for  $R \le \sqrt{2} H^{-1}$. The reference scale is slightly  shifted
 from $H^{-1}$ to $\sqrt{2}H^{-1}$ compared to $k=1$ case. However the connection between
 entropy bounds and scales is still transparent.

For the reheating after inflation, one finds from Fig.2 that $R >
H^{-1}$. In the case of $k=0$, the cosmological D-bound of $S \le S_{\rm D }$
 is  valid for $R \ge H^{-1} \sqrt{2+R^2/l^2_4}$,
 while the Bekenstein-Verlinde  bound of $S \le S_{\rm BV}$
 is valid for  $R \le H^{-1} \sqrt{2+R^2/l^2_4}$.
Here the cosmological D-entropy $S_{\rm D}$ is given by $\sqrt{S_{\rm H}^2-S_{\rm \Lambda}^2}$
with  $S_{\rm H} \ge S_{\rm \Lambda}$   from Eq.(\ref{2eq19}).  $S_{\rm H}$
is based on the Bekenstein-Hawking entropy of  Hubble-size black hole
($2 V_{\rm H}/4 G_4 H^{-1}=2 \pi H^{-2}/3 G_4 \equiv 2S_{\rm HS}$/3) times the number
($N_{\rm H}= V/V_{\rm H}$) of  Hubble regions in the universe,
 while $S_{\rm \Lambda}$
is based on the Bekenstein-Hawking entropy for de Sitter-event horizon
($2V_{\rm dS}/4 G_4 l_4=2 S_{\rm dS}/3 $) times the number
($N_{\rm dS}= V/V_{\rm dS}$) of de Sitter region in the universe.
From Eq.(\ref{2eq18}) and Fig.2 we have $l_4 \ge H^{-1}$.
Then one has $S_{\rm dS} \ge S_{\rm HS},~
V_{\rm dS} \ge V_{\rm H},~ N_{\rm H} \gg N_{\rm dS}.$ Hence one
achieves $S_{\rm H} \ge S_{\rm \Lambda}.$
Choosing $R \ge l_4$ during RH,
 then one finds the relation $R \ge l_4 \ge H^{-1}$. In this case
 the cosmological D-entropy bound  provides an  upper bound
 for RH, as the Hubble  bound did in RD.
 From Fig.2, the other case of $R \le H^{-1} \sqrt{2+R^2/l^2_4}$
seems to not occur in the reheating phase.

 Finally for the slow-roll period we approximate the entropy bound
 of the universe by the single de Sitter entropy : $S \le S_{\rm dS}$.

In conclusion, to estimate the entropy of the inflation we apply
three holographic entropy bounds to the three steps of SR, RH,and RD.
The relevant entropy bounds are the  de Sitter entropy bound for SR: $S \le S_{\rm dS}$,
the D-entropy bound for RH : $S \le S_{\rm D}$, and the Hubble entropy bound for RD :
$S \le S_{\rm H}$.  We
usually assume  the large increase of entropy via the reheating
process in the inflationary scenario.
In this work we just employ the entropy bounds for a
radiation-dominated universe with a positive cosmological  constant
for RH. We do not yet  have a detail mechanism to produce the
large entropy in a holographic way.

\section*{Acknowledgment}
We thank H. W. Lee and C.-R.Cai for helpful discussions.
This work  was supported in part by KOSEF, Project Number: R02-2002-000-00028-0.

\end{document}